# Breakdown of hawking evaporation opens new mass window for primordial black holes as dark matter candidate

Valentin Thoss ⍟,[1,2,3]★ Andreas Burkert[1,2,3] and Kazunori Kohri ⍟[4,5,6]

[1]*Universitäts-Sternwarte, Ludwig-Maximilians-Universität München, Scheinerstr. 1, D-81679 Munich, Germany*
[2]*Max-Planck Institute for Extraterrestrial Physics, Giessenbachstr. 1, D-85748 Garching, Germany*
[3]*Excellence Cluster ORIGINS, Boltzmannstrasse 2, D-85748 Garching, Germany*
[4]*Division of Science, National Astronomical Observatory of Japan (NAOJ), and SOKENDAI, 2-21-1 Osawa, Mitaka, Tokyo 181-8588, Japan*
[5]*Theory Center, IPNS, and QUP, KEK, 1-1 Oho, Tsukuba, Ibaraki 305-0801, Japan*
[6]*Kavli IPMU (WPI), University of Tokyo, Kashiwa, Chiba 277-8568, Japan*

## ABSTRACT
The energy injection through Hawking evaporation has been used to put strong constraints on primordial black holes as a dark matter candidate at masses below $10^{17}$ g. However, Hawking's semiclassical approximation breaks down at latest after half-decay. Beyond this point, the evaporation could be significantly suppressed, as was shown in recent work. In this study we review existing cosmological and astrophysical bounds on primordial black holes, taking this effect into account. We show that the constraints disappear completely for a reasonable range of parameters, which opens a new window below $10^{10}$ g for light primordial black holes as a dark matter candidate.

## 1 INTRODUCTION

The hypothesis of black holes forming in the early universe has been discussed for more than 50 years (Zel'dovich & Novikov 1967; Hawking 1971; Carr & Hawking 1974), with Chapline (1975) first to suggest that primordial black holes (PBHs) could constitute the entire dark matter of the universe. Since the 1970s, people have studied the consequences of PBHs as a dark matter candidate from the Planck mass $M_{\rm PBH} = M_{\rm pl}$ up to the' incredulity limit'[1] beyond $M_{\rm PBH} \sim 10^{10}\,M_\odot$. This has led to strong bounds that exclude PBHs of a single mass from constituting the entirety of the dark matter with the exception of a mass window in the asteroid range $M_{\rm PBH} \in [10^{17}, 10^{22}]$ g (Carr et al. 2021, and references therein).

The lower limit is a result of constraints due to black hole evaporation at low masses. This process was first described by Hawking (1974), as he was studying the consequences of light PBHs. He showed that a black hole will emit a thermal spectrum of particles, with the temperature of the radiation scaling as $T \sim 1/M_{\rm PBH}$. The described evaporation process is self-similar and ends with a final burst as $M \to 0$.

It was soon realized that the energy injection from low-mass PBHs is in conflict with observations of $\gamma$ rays, the cosmic microwave background (CMB) and the abundance of light elements produced during big bang nucleosynthesis (BBN) unless these black holes constitute only a tiny fraction of the dark matter (Chapline 1975; Hawking 1975; Novikov et al. 1979; Carr et al. 2010, for a historical overview). Furthermore, if the PBHs have a mass below $M \simeq 5 \times 10^{14}$ g, they would have completely evaporated by now [see Auffinger (2023) for a review on constraints of evaporating PBHs].

However, it is possible to avoid some of the constraints that are a result of black hole evaporation. Pacheco et al. (2023) have studied 'quasi-extremal' PBHs and found that they can be a viable dark matter candidate. Friedlander et al. (2022) and Anchordoqui, Antoniadis & Lüst (2022) have investigated PBHs in the context of large extra dimensions (Arkani-Hamed, Dimopoulos & Dvali 1998) and showed that this opens up new mass windows for light PBHs as dark matter candidate.

Even in the case of non-spinning, uncharged 4D black holes, it has always been clear that Hawking's semiclassical (SC) calculations will break down before the black hole vanishes entirely. Previously, this breakdown has been assumed to happen when the mass of the black hole reaches the Planck mass. While Hawking (1975) acknowledged this, he argued that the black hole will nevertheless completely disappear. Others have discussed the idea that the evaporation comes to a halt, leaving behind Planck-mass relics that can make up the entirety of the dark matter and avoid any constraints (MacGibbon 1987; Barrow, Copeland & Liddle 1992; Torres 2013; Taylor et al. 2024).

However, Dvali et al. (2020) have shown that the SC approximation will break down at a much earlier time – at latest when the black hole has lost roughly half of its initial mass. Hawking's result entirely neglects the backreaction of the emission on the quantum state of the black hole itself. However, this effect can no longer be ignored when the energy of the released quanta becomes comparable to that of the black hole. The crucial insight by Dvali et al. (2020) is that this backreaction leads to a universal effect of so-called 'memory burden', first introduced by Dvali (2018). This will significantly suppress further evaporation, which opens a possibility for light PBH to be a viable dark matter candidate as was already pointed out by Dvali et al. (2020).

★ E-mail: vthoss@mpe.mpg.de
[1] This term was coined by B. Carr and refers to the limit that at least one black hole must exist in a given environment (e.g. galaxy and universe).

In our work, we want to investigate the constraints on PBHs that are subject to the effect of memory burden and compare them to the results obtained within the SC Hawking picture. Following Dvali et al. (2020), we will describe the strength of the suppression by a single parameter $k$ and study the bounds on the dark matter fraction of PBHs $f_{\rm PBH}(k, M_{\rm PBH})$.

We want to emphasize that there is no precise understanding of the evaporation process beyond the SC regime yet. Therefore, our results should be understood as a rough guide on how the landscape of PBH constraints changes as one goes beyond the Hawking picture. In addition to the memory burden effect, there are also other potential quantum properties of black holes such as vortices (Dvali, Kühnel & Zantedeschi 2022; Dvali et al. 2023), which can affect the evaporation that we do not investigate here.

We want to mention the work of Alexandre, Dvali & Koutsangelas (2024), who studied the effect of memory burden on PBH constraints from BBN and CMB distortions, and her results are complimentary to our study. In addition, we want to point out the work of Dvali, Kühnel & Zantedeschi (2021), who discussed a new mass window for PBH due to memory burden for the case of $k = 2$ in the context of the PBH formation mechanism of quark confinement.

This paper is organized as follows: In Section 2, we discuss the memory burden effect, our modified model of the evaporation process, and how we compute the various constraints. Our results are presented and discussed in Section 3. We conclude with a summary in Section 4.

## 2 METHODS

In this section, we briefly introduce the physics of black hole evaporation and how it is modified in our model of the memory burden effect. We then discuss the methodology to obtain important constraints on PBHs. We conclude by reviewing further bounds that we do not study in this work in more detail.

Henceforth, we will denote the mass of PBHs at formation by $M_0$, and their mass today by $M_t$. To keep our results as general as possible and to make comparisons with previous results easier, we use a monochromatic mass function. Constraints for arbitrary mass distributions can be derived from our results. The quantity $\beta_{\rm PBH} = \rho_{\rm PBH, 0}/\rho_0$ gives the fraction of the universe's density in PBHs at the time of their formation. $f_{\rm PBH}$ refers to the present dark matter fraction of the PBHs, while $f_{\rm PBH, 0}$ denotes the dark matter fraction at formation time. They are related via

$$f_{\rm PBH} = f_{\rm PBH,0} \frac{M_t}{M_0}, \qquad f_{\rm PBH,0} = \frac{\beta_{\rm PBH}}{\Omega_{\rm DM}}, \qquad (1)$$

where $\Omega_{\rm DM}$ is the dark matter density at the time of PBH formation. We will use standard values for the Hubble rate $h = 0.67$, the number of relativistic degrees of freedom at formation $g_* = 106.75$[2] and for the factor $\gamma = 1$, which gives the fraction of mass inside a Hubble volume when the overdensity reenters the horizon that ends up in the black hole [see equations 2–6 of Carr et al. (2021)].

### 2.1 Semiclassical evaporation

The SC emission rate of a particle species $i$ with energy $E$ from a black hole with current mass $M$ and temperature $T$ is given by Hawking (1975).

$$\frac{{\rm d}^2 N_{i,{\rm SC}}}{{\rm d}E{\rm d}t}(E, M, s_i) = \frac{g_i}{2\pi\hbar} \frac{\Gamma(E, M, s_i)}{e^{E/k_B T(M)} - (-1)^{s_i}}, \qquad (2)$$

where $g_i$ specifies the degrees of freedom of the particle emitted, $s_i$ its spin, and $\Gamma$ are the greybody factors. The black hole temperature $T$ is directly related to its mass via

$$k_B T = \frac{\hbar c^3}{8\pi G M}. \qquad (3)$$

The emission of fundamental particles leads to a mass loss rate $\dot{M} = -\mathcal{F}(M)/M^2$ of the black hole, where $\mathcal{F}(M)$ encompasses the degrees of freedom of the particles that the black hole emits at a certain mass. The lifetime of a black hole with mass $M$ in the Hawking picture is given by $t_{\rm SC}(M) \approx M^3/(3\mathcal{F}(M))$.

Tabulated values for $\mathcal{F}(M)$ and the greybody factors are taken from BLACKHAWK (Arbey & Auffinger 2019). This is a publicly available code that is able to compute the SC emission rates for black holes with arbitrary mass- and spin-distribution. It makes use of existing particle physics codes to compute the secondary emission (see Section 2.2).

### 2.2 Secondary emission

Many of the particles that a sufficiently light black hole emits are not stable and decay or annihilate that leads to additional secondary emission of stable particles such as photons. The secondary emission of particles of type $i$ is given by

$$\frac{{\rm d}^2 N_{i,{\rm sec}}}{{\rm d}E{\rm d}t} = \sum_j \int {\rm d}E' \, {\rm Br}_{j \to i}(E, E') \frac{{\rm d}^2 N_j}{{\rm d}E'{\rm d}t}, \qquad (4)$$

where ${\rm Br}_{j \to i}(E, E')$ denotes the branching ratios that are calculated from particle physics codes. In our case, we use the spectra computed by BLACKHAWK and additionally make use of the code HDMSPECTRA (Bauer, Rodd & Webber 2021) for energies beyond the GeV-scale[3] BLACKHAWK makes use of the codes PYTHIA (Sjöstrand et al. 2015), HERWIG (Bellm et al. 2016), and HAZMA (Coogan, Morrison & Profumo 2020). All of them are suited for different energy ranges. For our purposes, we use HAZMA for black holes with $k_B T < 0.1$ GeV, HDMSPECTRA for temperatures $k_B T > 100$ GeV, and HERWIG in the intermediate range. Note that for a given black hole temperature, we include secondary emission only if $E > 10^{-6} k_B T$, as we do not trust the results at lower energies.

### 2.3 Memory burden

In the SC description of the evaporation process, a black hole decays self-similarly and can be fully described by its mass, spin, and charge without any knowledge about its prior history. During this process, the black hole entropy $S$, given by

$$S(M) = \frac{4\pi M^2 G k_B}{\hbar c} \approx 2.6 \times 10^{10} k_B \left(\frac{M}{1\,{\rm g}}\right)^2, \qquad (5)$$

---

[2]There are 28 bosonic and 90 fermionic degrees of freedom in the Standard Model giving $g_* = 28 + 7/8 \times 90 = 106.75$.

[3]By the time of publication, HDMSPECTRA has now been implemented into BLACKHAWK.

continually decreases. Since the outgoing Hawking radiation is purely thermal and contains no information, this leads to the well-known' paradox', as no information escapes the black hole but its storage capacity decreases. The memory burden effect circumvents this problem, as the information stored within the black hole stabilizes it against further decay. This effect is not a specific property of black holes but is a universal phenomenon shared among all quantum systems that have a maximum entropy.

When a black hole is initially formed, its quantum state is such that it has a high capacity of information storage. This is achieved by having a large number of so-called 'memory modes' that store the information and which are nearly gapless to achieve the high storage capacity. Dvali et al. (2020) have shown under very general assumptions that the decay of the black hole (i.e. the Hawking radiation) leads to a backreaction on the black hole itself that increases the energy gaps of the memory modes slowing down any further decay. The slowdown happens at latest when the black hole has lost on the order of half of its original mass (although it might happen much earlier as we will discuss later). At this point, the cumulative backreaction of all previously emitted particles becomes so strong that the SC approximation breaks down.

As Dvali et al. (2020) have discussed, there are two possibilities for the fate of an evaporating black hole beyond half-decay. Either it continues to emit quanta with a strongly suppressed rate due to the memory burden effect or a new classical instability sets in. In the latter case, light PBHs cannot constitute the dark matter. However, if the evaporation is strongly suppressed, there is the possibility that PBHs can be a viable dark matter candidate. This is the scenario that we are investigating here.

### 2.4 Modified evaporation

Our modified model of the evaporation process assumes the validity of Hawking's results until the black hole has reached a mass of $M = qM_0$. Unless otherwise stated, we conservatively use $q = 1/2$ following the previous discussion. Beyond this point, the emission rate is suppressed (denoted by MB for memory burden), which we will parametrize following Dvali et al. (2020) as

$$\frac{d^2 N_{i,\mathrm{MB}}}{dE dt}(E, M_0, s_i) = \frac{1}{(S(qM_0)/k_B)^k} \frac{d^2 N_{i,\mathrm{SC}}}{dE dt}(E, qM_0, s_i), \quad (6)$$

where $k$ is an exponent that controls the strength of the suppression and $S$ is the black hole entropy [equation (5)].

Note that the large value of the entropy implies that the emission rate decreases by many orders of magnitude for $k \sim 1$. In this model, the evaporation rate and the black hole temperature remain constant. This implies a linear decay of the black hole with lifetime $t_{\mathrm{MB}}(M) \approx \frac{(qM_0)^3 S_0^k}{\mathcal{F}(qM_0)}$, neglecting the time spent in the SC regime.

As an alternative scenario, we could let the black hole temperature increase in the usual way, leading to a black hole explosion and

$$\frac{d^2 N_{i,\mathrm{MB}}}{dE dt}(E, M, s_i) = \frac{1}{S(M)^k} \frac{d^2 N_{i,\mathrm{SC}}}{dE dt}(E, M, s_i), \quad (7)$$

This leads to a very similar lifetime $t_{\mathrm{MB}}(M) \approx \frac{(qM_0)^3 S_0^k}{(3+2k)\mathcal{F}(qM_0)}$ and only makes a meaningful difference to the other scenario when the lifetime of the black hole is comparable to or less than the Hubble time. However, such scenarios are heavily constrained ($f_{\mathrm{PBH},0} \ll 1$) as we will show in this work. For simplicity, and since we do not know the late-time behaviour of the black hole in such detail, we will restrict our discussion to the former scenario. We want to emphasize that (in both cases) due to the dependence on the initial black hole mass $M_0$, the evaporation process is no longer self-similar.

### 2.5 Galactic $\gamma$ ray emission

If light PBHs make up a sizeable fraction $f_{\mathrm{PBH}}$ of the dark matter halo of the Milky way, they would be detectable through their photon emission. Comparisons with observations have been used to put strong constraints on PBHs in the mass range $M_0 \in [5 \times 10^{14}, 10^{17}]$ g [e.g. Carr et al. (2016), see also Auffinger (2023) for a recent review].

The measured flux of photons would be

$$\Phi_{\mathrm{PBH}} = \frac{f_{\mathrm{PBH}}}{4\pi M_t \Delta\Omega} \frac{d^2 N_\gamma}{dE dt} \int_{\Delta\Omega} d\Omega \int dr \rho_{\mathrm{DM}}(R(r, l, b)), \quad (8)$$

with $\Delta\Omega$ being the observed field of view on the sky and $R$, $l$, and $b$ the Galactocentric distance, longitude, and latitude, respectively. The constraint on $f_{\mathrm{PBH}}$ is obtained by requiring that

$$\int_{E_{\mathrm{low}}}^{E_{\mathrm{up}}} dE\, \Phi_{\mathrm{PBH}} \leq \Phi_{\mathrm{gc}}(E_{\mathrm{up}} - E_{\mathrm{low}}), \quad (9)$$

where $\Phi_{\mathrm{gc}}$ is the measured $\gamma$ ray flux in the energy bin $[E_{\mathrm{low}}, E_{\mathrm{up}}]$.

For $\rho_{\mathrm{DM}}(R)$, we choose an NFW profile with the 'convenient' set of parameters from McMillan (2011). This means our results will be identical to Auffinger (2022) in the SC limit, since we also use the observational data from their ISATIS code. This includes data from *INTEGRAL*, *COMPTEL*, *EGRET*, and *Fermi-LAT* (Strong et al. 1994; Strong & Mattox 1996; Strong et al. 1999; Bouchet et al. 2011; Strong 2011). Since we are also interested in PBHs of lower masses (and thus higher spectral energies) than usually investigated, we extend this data set by the recent observational data from *LHAASO* in the energy band from $10^4$ to $10^6$ GeV (Cao et al. 2023, 'inner region').

### 2.6 Extragalactic $\gamma$ ray background

The evaporation of PBHs between the time of recombination $t_{\mathrm{rec}}$ and today $t_0$ will lead to a contribution to the extragalactic $\gamma$ ray background. This has been used to constrain PBHs with masses $M_0 \in [3 \times 10^{13}, 10^{17}]$ g (Carr et al. 2010; Arbey, Auffinger & Silk 2020; Ballesteros, Coronado-Blázquez & Gaggero 2020; Chen, Zhang & Long 2022).

Assuming a homogenous distribution of dark matter, the observed flux is

$$\Phi_{\mathrm{PBH}} = \frac{c n_t}{4\pi} \int_{t_{\mathrm{rec}}}^{t_0} dt\, (1+z) \frac{d^2 N_\gamma}{dE dt}\left((1+z)E, M(t)\right), \quad (10)$$

where $n_t$ is today's number density of the PBHs which is related to $f_{\mathrm{PBH},0}$ via

$$n_t \approx 2.2 \times 10^{-30}\,\mathrm{cm}^{-3}\, f_{\mathrm{PBH},0} \left(\frac{M_0}{1\mathrm{g}}\right)^{-1}. \quad (11)$$

The constraint on $f_{\mathrm{PBH},0}$ (and $f_{\mathrm{PBH}}$) is obtained analogously to equation (9). We again use the observational data available in the ISATIS code from *HEAO*, *COMPTEL*, *EGRET*, and *Fermi-LAT* (Gruber et al. 1999; Strong, Moskalenko & Reimer 2004; Ackermann et al. 2015; Ruiz-Lapuente et al. 2016). In addition, we include the flux data from *LHAASO* (Cao et al. 2023, 'outer region').

### 2.7 CMB anisotropies

The injection of energy from PBH evaporation after recombination can ionize the otherwise neutral medium. This in turn leads to the rescattering of CMB photons that affects the angular power spectrum of temperature and polarization. Comparisons with measurements of the CMB have led to strong constraints on PBH in the mass range $M_0 \in [3 \times 10^{13}, 10^{17}]$ g (Poulin, Lesgourgues & Serpico 2017; Stöcker et al. 2018; Acharya & Khatri 2020; Cang, Gao & Ma 2022). Stöcker et al. (2018) have developed the publicly available code EXOCLASS, a branch of the Boltzmann code CLASS (Blas, Lesgourgues & Tram 2011). It is able to compute the CMB power spectra for any value of $M_0$ and $f_{\text{PBH},0}$. To achieve this, the code computes the rate of energy density deposition as a function of redshift,

$$\frac{d^2 E}{dt dV}\bigg|_{\text{dep},\alpha}(z) = h_\alpha(z) \frac{d^2 E}{dt dV}\bigg|_{\text{inj}}(z) = \frac{h_\alpha(z) f_{\text{PBH},0} \rho_{\text{DM},t} (1+z)^3 \dot{M}}{M_0}, \quad (12)$$

where $\alpha$ denotes heating, ionization, or excitation+ionization as the possible channels. The energy deposition function $h_\alpha(z)$ is a convolution of the PBH spectrum with a transfer function $T^i_\alpha(z', z, E)$. This function gives the fraction of the energy $E$ injected at redshift $z'$ in a channel $\alpha$ that is deposited at redshift $z$ for the particle type $i$.

To obtain constraints on $f_{\text{PBH},0}$, we use EXOCLASS and MONTEPYTHON (Audren et al. 2013; Brinckmann & Lesgourgues 2019) to perform a Markov chain Monte Carlo (MCMC) analysis. The model uses the Planck TT,TE,EE+lowE+lensing likelihood (Planck Collaboration 2020) with $f_{\text{PBH},0}$ as an additional cosmological parameter. To perform the calculations, we modified EXOCLASS and implemented our model of the evaporation process as described in Section 2.4. In addition, we also change the computation of the PBH spectra in the DARKAGES module of the code. We replace the geometric optics approximation with the full results from BLACK-HAWK including secondary emission as described in Section 2.2. Note that in the SC limit, we reproduce the results from Stöcker et al. (2018) within a factor of a few.

The transfer function used in EXOCLASS is only tabulated up to $E \sim 6$ TeV and for higher energies the code simply uses the last tabulated value. This should be a reasonable approximation even for much higher energies as the only relevant cooling process at energies above $E \sim 1$ TeV is the pair production on the CMB and subsequent cascade to lower energy photons that produces a universal spectrum (Slatyer, Padmanabhan & Finkbeiner 2009). The final deposition efficiency does not depend on the initial photon energy and the value of the transfer function should thus remain constant with respect to the energy. Indeed, the tabulated values of the transfer function change very little for energies $E > 1$ TeV.

Since we have $k$ as an additional free parameter, it would be computationally very expensive to perform a large number of MCMC runs over the entire parameter space of $M_0$ and $k$ and given the simple nature of our model we do not aim for very precise results. In fact, as we are going to show in this work, the constraints from CMB anisotropies are weaker than those from galactic and extragalactic $\gamma$ rays by several orders of magnitude for most of the parameter space and thus not as relevant, which justifies approximations.

If the initial black hole mass is larger than $M_0 \simeq 3 \times 10^{13}$ g, then the SC phase of the evaporation ($M_0 \rightarrow qM_0$) is still ongoing after recombination. In this case, one can expect the constraints to change only by a factor of a few if $q \sim 1/2$, as the black hole still releases a sizeable fraction $1 - q$ of its total energy $M_0 c^2$. To investigate this, we perform a MCMC run for $M_0 = 5 \times 10^{13}$ g were the black hole evaporation is stopped after half-decay. We compare it to a simulation with the full evaporation process. The resulting constraints on $f_{\text{PBH},0}$ only differ by roughly 20 per cent. As this is significantly smaller than the modelling uncertainty of around two dex [see e.g. fig. 6 of Auffinger (2023)], we ignore the effect of memory burden in this mass range and just use the SC results from Stöcker et al. (2018).

To obtain the constraints for PBHs with $M_0 < 3 \times 10^{13}$ g, we run a full MCMC for $M_0 \in [10^3, 10^5, 10^7, 10^9, 10^{11}, 10^{13}]$ g and interpolate the results for other values of $M_0$ and $k$. To do this, we assume that $f_{\text{PBH}}$ scales linearly with the total amount of energy injection, as is motivated by equation (12). Effectively, we are rescaling the constraints on $f_{\text{PBH},0}(M_0, k)$ computed using EXOCLASS to obtain $f_{\text{PBH},0}(M'_0, k')$ through

$$f_{\text{PBH},0}(M'_0, k') \frac{\Delta M'}{M'_0} = f_{\text{PBH},0}(M_0, k) \frac{\Delta M}{M_0}, \quad (13)$$

where $\Delta M$ is the change in mass from recombination to the end of reionization which depends on $k$. This does not require computing an angular power spectrum and is treating $h_\alpha(z)$ as constant. The latter approximation is exact if we keep $M_0$ constant as $h_\alpha(z)$ only depends on the shape of the PBH spectrum, which is unaffected by $k$. We discuss this approximation, its computation, and its validity in more detail in Appendix B.

### 2.8 Big bang nucleosynthesis

The abundance of light elements in the universe is the earliest cosmological probe to study PBH evaporation. In the Hawking picture, black holes with masses $M_0 \in [10^{10}, 10^{13}]$ g evaporate during or after the formation of light elements. The emitted radiation will alter the neutron-to-proton ratio and lead to photo- and hadrodissociation of elements. Since the standard BBN scenario predicts the abundance of light elements with great success, any modification will be heavily constrained. Carr et al. (2021, updated from Carr et al. 2010) through results of Kawasaki et al. (2018) and Hasegawa et al. (2019)) find that the initial dark matter fraction is constrained to $f_{\text{PBH},0} \lesssim 10^{-3}$ in the denoted mass range with other studies obtaining comparable results (Acharya & Khatri 2020; Keith et al. 2020).

To understand how these results will be affected by the memory burden, we need to distinguish two regimes as in Section 2.7. For $M_0 \gtrsim 10^{10}$ g, the SC evaporation phase is still ongoing during or after nucleosynthesis. Therefore, one expects the constraints to change only by a factor of order unity as the black hole still releases half of its total energy $M_0 c^2$. To understand this in more detail we use of the model described in Kawasaki et al. (2018) and modify it by halting the evaporation process at half-mass. This is a good approximation for $k \gtrsim 0.5$, since the lifetime of the black hole exceeds the age of the universe for $M_0 > 10^{10}$ g in this case.

On the other hand, if the initial mass of the PBHs is below $M_0 \sim 10^{10}$ g then the memory burden slows down the evaporation before they can affect the abundance of light elements. If the black holes are still present today, then their evaporation rate in the early universe will be completely negligible (compare this to the SC case where there are no BBN bounds on PBHs with a lifetime greater than $t \sim 10^6$ yr). Conversely, if they are light enough to evaporate significantly in the early universe and affect the abundance of light elements then they will not survive until today. For our purposes, we can therefore ignore this regime. This is also justified by the results of Alexandre et al. (2024), who discuss the constraints from BBN in more detail.

### 2.9 Seismic constraints

If PBH are light enough and make up a sizeable fraction of the dark matter, then they will frequently come close to or even transit through the earth. The expected collision rate is (Luo et al. 2012)

$$\Gamma = 10^8 \, \text{yr}^{-1} f_{\text{PBH}} \left(\frac{M}{1 \, \text{g}}\right)^{-1} \left(\frac{\langle v \rangle}{200 \, \text{km/s}}\right) \quad (14)$$

where $\langle v \rangle$ is the mean dark matter velocity. This assumes that the PBHs are smoothly distributed throughout the solar system. While this would imply a sizeable number of collisions per year for $M \ll 10^8$ g, these are virtually unobservable. In the same work, it was found that a black hole with $M_0 = 10^{15}$ g would produce a seismic event with a magnitude of $M_w = 4$. Since the amplitude of the seismic waves scales linearly with the black hole mass according to Luo et al. (2012), this would imply immeasurable magnitudes $M_w \ll 1$ for black holes with mass $M_0 \ll 10^{12}$ g. While close encounters with the Earth would be more numerous, they are still too weak to be detectable in the mass range that we are interested in.

## 2.10 Other constraints

In addition to the constraints already discussed, there are a large number of other bounds on evaporating PBHs (for a full overview see the review of Auffinger 2023).

Many of these are conceptually similar to the constraints from the galactic $\gamma$ ray emission but instead look at the flux of electrons, positrons, or neutrinos (Boudaud & Cirelli 2019; Dasgupta, Laha & Ray 2020), the 511 keV annihilation signal (DeRocco & Graham 2019; Laha 2019), radio emission from synchrotron radiation Chan & Lee (2020), heating of the ISM (Kim 2021; Laha, Lu & Takhistov 2021) or take into account other astrophysical contributions to get stronger bounds (Berteaud et al. 2022). They are all usually within one order of magnitude of the conservative approach that we use here.

Another important constraint arises from spectral distortions of the CMB (Acharya & Khatri 2020; Chluba, Ravenni & Acharya 2020; Lucca et al. 2020). It affects black holes with masses in the range $M_0 \in [10^{11}, 10^{13}]$ g. One can follow the same arguments as in the previous section to conclude that the bounds will not change dramatically in the regime where the SC evaporation is still (partly) ongoing. During the phase of slowed down evaporation, the constraints only affect black holes which would not survive until today. This is also discussed in the work of Alexandre et al. (2024). Since the constraints from CMB spectral distortions are (currently) less strong than those from BBN and cover a smaller range of masses, we will ignore them for our purposes.

PBHs with a mass of $M_0 < 10^9$ g were previously ruled out as a dark matter candidate unless they could form stable relics at the Planck mass (MacGibbon 1987; Barrow et al. 1992; Torres 2013; Taylor et al. 2024). Nevertheless, PBHs in this mass range have been studied as an explanation for baryogenesis, as a source of particle dark matter and gravitational waves as well as a solution for the Hubble tension (Hawking 1975; Dolgov, Naselsky & Novikov 2000; Baumann, Steinhardt & Turok 2007; Fujita et al. 2014; Allahverdi, Dent & Osinski 2018; Lennon et al. 2018; Hooper, Krnjaic & McDermott 2019; Morrison, Profumo & Yu 2019; Baldes et al. 2020; Hooper et al. 2020; Masina 2021; Papanikolaou, Vennin & Langlois 2021; Papanikolaou 2023). All of these results need to be revisted in the context of the memory burden effect. In general, any regions of the parameter space for which $q\beta_{\text{PBH}} > \Omega_{\text{DM}}$ (i.e. $qf_{\text{PBH},0} > 1$) are not allowed in our scenario as it would overclose the universe.

When PBHs form in the early radiation-dominated universe, they are able to accrete significantly for a short period of time, thereby increasing their mass by a factor of a few (Escrivà 2022). However, this will have no meaningful effect on our results as the evaporation will start to become dominant long before the black hole reaches the memory burden stage. This can be seen by a rough comparison of the evaporation rate to the accretion rate $\dot{M} = f c \rho_r 4\pi r_{\text{PBH}}^2$, where $f$ is a factor of order unity and $\rho_r = 3/32\pi G t^2$ the radiation density in the early universe. Both rates are approximately equal at

$$t = \left(\frac{3GM_0^4}{2c^3 \mathcal{F}(M_0)}\right)^{1/2} \approx 5 \times 10^{-6} \tau_{\text{SC}}(M_0) \left(\frac{M}{1 \, \text{g}}\right)^{-1}, \quad (15)$$

where we have used that $\mathcal{F}(M_0)$ becomes constant for $M_0 < 10^{11}$ g. Therefore, for black holes much larger than the Planck mass, the duration of the accretion phase will be completely negligible to the SC duration of the evaporation phase $\tau_{\text{SC}}$. Our results should thus not be affected by accretion as long as $M_0$ is interpreted as the mass of the black hole at the end of the accretion phase (as is usually done in the literature).

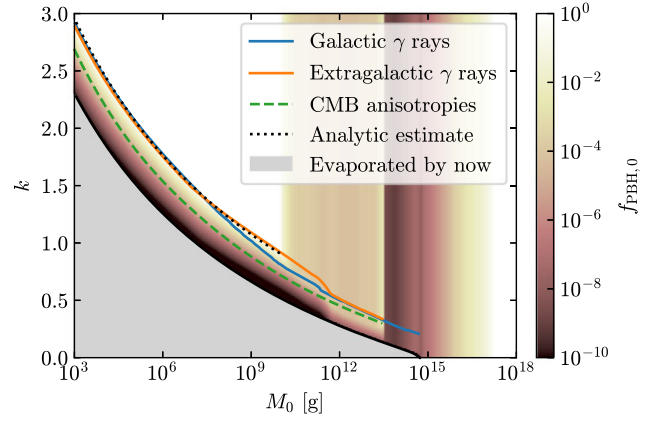

**Figure 1.** Combined constraints on $f_{\text{PBH},0}(k, M_0)$. The coloured lines show $f_{\text{PBH},0} = 1$ contours for each type of constraint. In the white region, PBHs can make up the entirety of the dark matter. The black, dotted line displays the analytic estimate given by equation (16). The different linestyles are chosen for better readability.

Finally, we want to mention that in the standard formation scenario of PBHs from spherical collapse of overdensities, there is a lower mass limit of $M_0 \sim 0.1$ g that is set by the time of the end of inflation.

## 3 RESULTS AND DISCUSSION

In this section, we discuss the results and possible uncertainties in our model. The combined map of all constraints is shown in Fig. 1. The results for galactic and extragalactic $\gamma$ ray emission are also shown separately in the appendix.

### 3.1 Galactic $\gamma$ ray emission

Fig. A1 shows the constraints obtained from galactic $\gamma$ ray emission. PBHs with initial mass $M_0 \gtrsim 5 \times 10^{14}$ g have not yet lost half of their mass and thus the usual constraints apply which rule out PBHs to be the entirety of the dark matter up to $M_0 \sim 10^{17}$ g independent of $k$. For $k > 0.2$, a new mass window emerges that extends down to $M_0 = 10^9$ g for $k = 1$ and to $M_0 = 4 \times 10^5$ g for $k = 2$. Note that when only primary photon emission is considered (dashed lines in Fig. A1), then there are no bounds for $k > 1.3$ or $M < 6 \times 10^5$ g beyond the trivial constraint that they would have evaporated by now. However, when considering the emission of secondary photons, we obtain bounds on these PBHs far beyond these limits. We do not find any constraints from galactic $\gamma$ ray emission for $M_0 < 3 \times 10^2$ g.

The radiation of these light black holes goes beyond the energy range covered by any current $\gamma$ ray observations. In this case, one would need to consider indirect effects of these highly energetic photons (beyond PeV) to obtain limits. For each value of $k$, the constraints on $f_{\mathrm{PBH},0}$ extend roughly two orders of magnitude beyond the mass at which the PBHs would completely evaporate by today.

### 3.2 Extragalactic $\gamma$ ray background

In Fig. A2, the constraints from the extragalactic $\gamma$ ray background are shown. In the Hawking picture, the dark matter fraction of PBHs is limited to $f_{\mathrm{PBH},0} < 1$ in the mass range from $M_0 \approx 3 \times 10^{13}$ g to $M \sim 10^{17}$ g. As explained before, for initial black hole masses $M \gtrsim 5 \times 10^{14}$ g, there are no changes. In the range $M_0 \in [3 \times 10^{13}, 5 \times 10^{14}]$ g, the constraints are softened at most by one order of magnitude when taking into account all observational data. This is because the SC evaporation phase is still ongoing after recombination. The lower end of the constrained mass range (which is roughly set by the mass of a black hole that evaporates at recombination) changes by only 4 per cent. A new window for PBHs as dark matter candidates opens at lower masses for $k > 0.35$. It extends down to $M = 10^9$ g for $k = 1$ and to $M = 10^5$ g for $k = 2$. If only primary photon emission is considered, then the constraints become significantly weaker as can be seen from the dashed lines in Fig. A2. We are not able to constrain black hole holes with $M_0 \lesssim 10^2$ g as the energy exceeds the current observational range, as already discussed for the galactic $\gamma$ ray emission.

### 3.3 Other constraints

The constraints on PBHs from CMB anisotropies are weaker than those from galactic and extragalactic $\gamma$ ray emission by 4–6 orders of magnitude for $M_0 < 10^{12}$ g and by around 2 dex for $M_0 \in [10^{12}, 3 \times 10^{13}]$ g. Note that for $M_0 < 10^9$ g, our results rely on extrapolation of the transfer function and are thus subject to some uncertainty. For $M_0 > 3 \times 10^{13}$ g, we just give the constraints from Stöcker et al. (2018), as explained in Section 2.7.

The bound from BBN in the mass range $M_0 \in [10^{10}, 10^{13}]$ g weakens by around 45 per cent if the evaporation is stopped at half-mass. As discussed in Section 2.8, there are no other constraints from BBN for black holes that survive to the present day.

### 3.4 Combined constraints

The combined map of constraints $f_{\mathrm{PBH},0}(k, M_0)$ is shown in Fig. 1. Fig. 2 shows $f_{\mathrm{PBH},0}(M_0)$ for $k = 2$, which is chosen for illustrative purposes. One can identify the constraints from the SC evaporation phase as the vertical, almost $k$-independent band in the range $M_0 \in [10^{10}, 10^{17}]$ g. The slowed down evaporation phase leads to constraints that are dependent on $k$. For $k \gtrsim 1.0$, a new mass window emerges for which PBHs can make up the entirety of the dark matter. The upper mass limit of $M_0 \sim 10^{10}$ g is set by the time of nucleosynthesis as the PBHs must reach the stage of suppressed evaporation before the onset of BBN. On the other hand, the lower bound depends on the strength of the suppression $k$, as it determines the lifetime of the black hole.

We can understand our results in simple terms by considering the fraction $\alpha$ of the PBH mass that evaporates during memory burden until today ($qM_0 \to (1-\alpha)qM_0$). From equation (6), it follows that

$$M_0 \approx \left(\frac{\mathcal{F}(qM_0)t_0}{\alpha}\right)^{\frac{1}{3+2k}}\left(\frac{\hbar c}{4\pi G}\right)^{\frac{k}{3+2k}}, \quad (16)$$

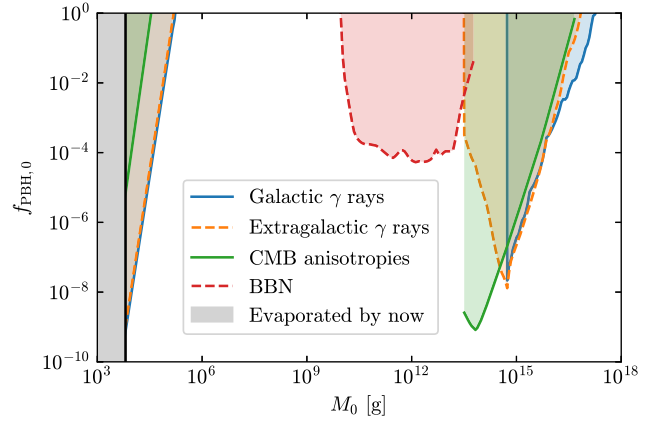

**Figure 2.** Constraints on $f_{\mathrm{PBH},0}(M_0)$ for $k = 2$. The different linestyles are chosen for better readability.

where $t_0$ is the age of the universe. This is is plotted in Fig. 1 for $\alpha = 10^{-13}$ and $\mathcal{F}(qM_0) \approx \mathcal{F}(0) \approx 8.2 \times 10^{26}$ g$^3$ s$^{-1}$ and gives a good approximation for the lower bound on the black hole mass in the parameter range that we study. Put differently, this implies that the memory burden effect must increase the lifetime of the black hole to at least $\sim 10^{13}$ times the age of the universe in order to avoid all existing constraints.

### 3.5 Model uncertainties

Due to the lack of a detailed understanding of the evaporation process beyond the SC limit, we keep our model of the evaporation process as simple as possible. Motivated by Dvali et al. (2020), we model the strength of the suppression by a factor $1/S_0^k$ with $S_0$ being the black hole entropy. This means $k$ only depends logarithmically on other parameters in our models. Therefore, any uncertainties in our models of the constraints will only weakly affect the bound on $k(M_0)$, where $f_{\mathrm{PBH},0} = 1$ (i.e. the coloured lines in Fig. 1). More specifically, if e.g. the rate of galactic $\gamma$ ray emission decreases by a factor of 10, this will shift the bound $k(M_0)$ by less than 0.05 for the range of mass that we study.

Another uncertainty in our model is the question on how to set the black hole temperature and its spectrum. When the SC approximation breaks down, we can no longer expect the emission to be thermal. In our model, we keep the shape of the spectrum fixed as a zeroth-order approximation. Should the spectrum change very drastically (but keeping d$N$/d$t$ constant) then this will most strongly affect the constraints from galactic and extragalactic emission. If the emission becomes more soft, then the bounds would become weaker and vice versa. This is due to the larger observed $\gamma$ ray flux at lower energies. However, even a drastic change in $f_{\mathrm{PBH}}$ will not shift the bound on $k(M_0)$ by much as explained. The constraints from CMB anisotropies are very robust to uncertainties in the black hole spectrum since they depend mostly on the total rate of emission. In fact, in the mass range that we study, the transfer function is approximately constant (see Section 2.7) w.r.t. to the energy.

Similar arguments can be made about the uncertainties regarding the secondary emission of the PBHs. During the phase of memory burden, it is in principle possible that the emission of particles is no longer 'democratic', i.e. the emission of certain particle species are preferred [see also the discussion in Alexandre et al. (2024)]. However, as long as the emission rate only changes within an order of magnitude, our bounds should not change dramatically.

Regarding the black hole temperature, we discussed two different approaches in Section 2.4. For the results presented so far, we keep it fixed once the black hole has lost half of its mass. If we instead assume $T \sim 1/M$ throughout the memory burden phase [i.e. as in equation (7)] then we find that it changes our results only when the lifetime is comparable to or less than a Hubble time. Thus, it only affects the constraints in a regime, where PBHs are already excluded from being a sizeable fraction of the dark matter ($f_{PBH,0} \sim 10^{-10}$).

Finally, for our analysis, we defined the onset of the memory burden effect when the black hole loses half of its initial mass by setting $q = 1/2$. As Dvali et al. (2020) have shown, this is the latest time at which a quantum backreaction is unavoidable. Dvali & Panchenko (2015) and Michel & Zell (2023) have shown that in certain prototype systems of black holes, the breakdown of Hawking's calculations could happen already after a fraction $1/\sqrt{S}$ of the SC lifetime. While this requires more microscopic justification, it is still worthwhile to consider the effect on the constraints of PBH. Such an early transition into the phase of memory burden would correspond to

$$1 - q \approx 2 \times 10^{-6} (M_0/\text{g})^{-1}, \quad (17)$$

which would imply that black holes with $M_0 > 10^{10}$ g reach the memory burden phase after losing a fraction of less than $10^{-15}$ of their initial mass. This implies that any constraints that arise from the SC evaporation phase (the 'vertical band' in Fig. 1) will be significantly weakened. When we recompute these constraints under this assumption, then PBHs can make up the entirety of the dark matter for $M_0 > 10^{10}$ g. One is still left with the constraints from the slowed down phase of the evaporation. This would open a significantly larger window for PBHs as a viable dark matter candidate.

## 4 SUMMARY AND CONCLUSIONS

In this work, we investigate how the constraints on evaporating PBHs change when one goes beyond the SC calculations from Hawking (1974). When the quantum backreaction of the emitted particles is taken into account, the evaporation can slow down drastically due to the effect of memory burden described by Dvali et al. (2020). Here, we compute the bounds on the dark matter fraction of PBHs as a function of the currently unknown parameter $k$ that quantifies the strength of the memory burden effect. It gives the suppression of the evaporation rate in powers of the black hole entropy $1/S^k$.

We find that for $k > 1.0$ a new mass window emerges, where PBHs can be a viable dark matter candidate. It extends up to $M_0 \sim 10^{10}$ g with higher mass being ruled out due to constraints from BBN. The lower end of the mass window depends on the parameter $k$, extending to $M_0 \sim 10^5$ g for $k = 2$.

The bounds that we obtain with our simple model of the evaporation process are quite robust and should hold unless the breakdown of the SC evaporation phase happens much earlier or the spectra of the PBH evaporation changes dramatically. It should be noted that all the constraints that we obtain in this work are computed for a monochromatic mass function and zero spin of the black hole to keep our results as general as possible. In reality, PBHs will form with an extended distribution of initial mass. Studying bounds on the dark matter fraction of PBHs for more realistic mass functions and a non-zero spin distribution is left for future work.

Our results are supposed to provide a first insight on the landscape of PBH constraints beyond the Hawking picture of black hole evaporation. A more detailed understanding of the actual evaporation process remains a significant theoretical challenge but will undoubtedly help to study the possibility that dark matter exists in the form of light PBHs.

In addition to open questions about the fundamental nature of black holes, it would also be of great interest to study the potential observational signatures and possibilities to detect PBHs in this new mass window where they are a viable dark matter candidate (see e.g. Lehmann et al. 2019).

## SOFTWARE

## ACKNOWLEDGEMENTS

We thank Gia Dvali for giving us the idea for this project, as well as for his supervision. We also would like to thank Ana Alexandre, Manuel Behrendt, Florian Kühnel, Tracy Slatyer, and Sebastian Zell for fruitful discussion and helpful comments. We are grateful for the help of Patrick Stöcker and Vivian Poulin with regards to the EXOCLASS code. This research was supported by the Excellence Cluster ORIGINS, which is funded by the Deutsche Forschungsgemeinschaft (DFG, German Research Foundation) under Germany's Excellence Strategy—EXC-2094 - 390783311. Kazunori Kohri is supported by KAKENHI grant no. JP22H05270.

## DATA AVAILABILITY

The numerical results presented in this article are publicly accessible under https://doi.org/10.5281/zenodo.10718218 (Thoss 2024).

## APPENDIX A: ADDITIONAL FIGURES

Here we provide additional plots that show the constraints on $f_{\rm PBH, 0}(k, M_0)$ for galactic $\gamma$ ray emission in Fig. A1 and for extragalactic $\gamma$ ray emission in Fig. A2.

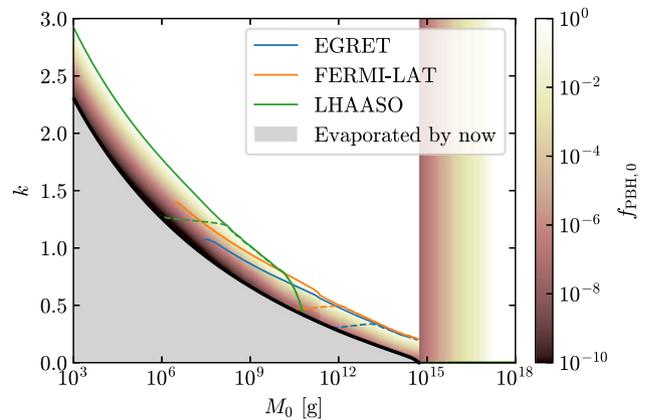

**Figure A1.** Combined constraints on $f_{\rm PBH, 0}(k, M_0)$ from galactic $\gamma$ ray emission. The coloured lines show $f_{\rm PBH} = 1$ contours separately for selected observational data. The dashed lines are computed using only the primary photon emission.

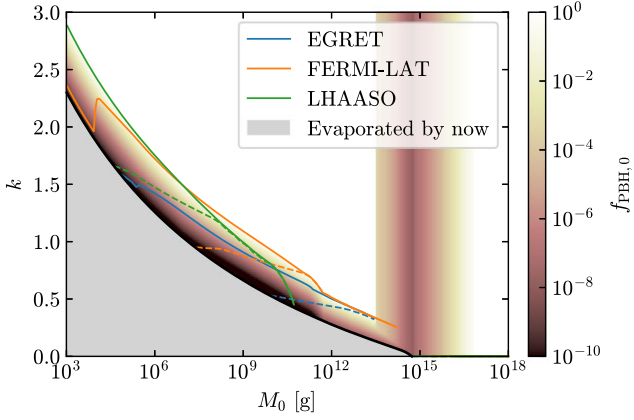

**Figure A2.** Combined constraints on $f_{\rm PBH,\,0}(k, M_0)$ from the extragalactic $\gamma$ ray background. The coloured lines show $f_{\rm PBH} = 1$ contours separately for selected observational data. The dashed lines are computed using only primary photon emission.

## APPENDIX B: CMB ANISOTROPIES

As explained in Section 2.7, to compute the constraints from CMB anisotropies, we rescale the results from the full MCMC runs using equation (13). In Fig. B1, we show that this method is able to reproduce the SC constraints within one order of magnitude by rescaling a single value $f(M_0 = 2.3 \times 10^{16}\text{g}) = 5.1 \times 10^{-2}$ obtained by Stöcker et al. (2018) over three orders of magnitude in mass. Note that in Fig. B1, we also show the resulting constraints when properly taking into account the $(1 + z)^3$ factor in equation (12). There is no noticeable difference down to $M_0 \sim 3 \times 10^{14}$g. The divergence at low masses is not relevant for us since it will in any case only affect those PBHs that will not survive until today – both in the SC limit as well as for $k > 0$.

To obtain constraints for $k > 0$, we compute six full MCMC runs for $M_0 \in [10^3, 10^5, 10^7, 10^9, 10^{11}, 10^{13}]$ g, where $k$ is chosen for each mass such that $f(k, M_0) = 10^{-2}$ when rescaled from $f(M_0$ $= 2.3 \times 10^{16}\text{g}) = 5.1 \times 10^{-2}$ to the respective masses according to equation (13) and taking into account the effect of the memory burden. The values of $f(k, M_0)$ obtained from the MCMC runs lie between $3.8 \times 10^{-2}$ and $9.7 \times 10^{-2}$ and thus up to one order of magnitude higher then the $10^{-2}$ is expected from rescaling. These results are then used to obtain constraints for arbitrary values of $k'$ and $M_0'$ by rescaling from the two nearest values of $M_0$ and interpolating between the results. To test the validity of this approach, we run a second set of MCMC runs with $M_0 \in [10^4, 10^6, 10^8, 10^{10}, 10^{12}]$ g and $k$ such that our method would give $f(k, M_0) = 10^{-1}$. The resulting constraints on $f_{\rm PBH,\,0}$ from the full computation are within a factor of two from our approximation, which is more than sufficient. This translates to an error in the bound on $k(M_0, f = 1)$ of less than 0.01.

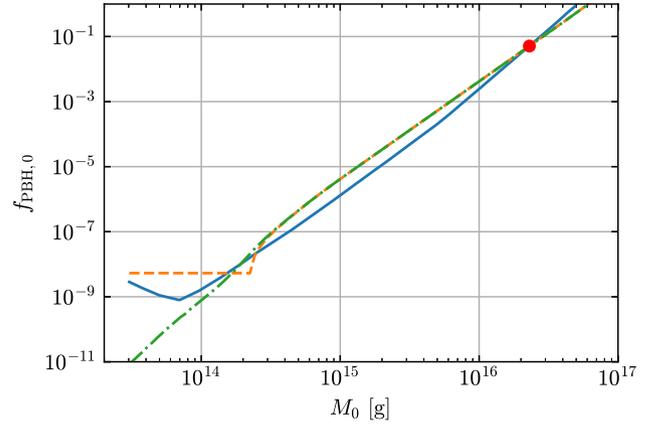

**Figure B1.** CMB constraints on the initial dark matter fraction of PBHs $f_{\rm PBH,\,0}(M_0)$ in the SC limit ($k = 0$). The solid blue line shows the result from Stöcker et al. (2018). The red dot shows the value $f(M_0 = 2.3 \times 10^{16}\text{g}) = 5.1 \times 10^{-2}$ from which the rescaled constraints were obtained. The dashed orange line shows the constraints obtained through equation (13), whereas the green dash–dotted line also takes into account the redshift-dependence of the integral.

This paper has been typeset from a T<sub>E</sub>X/L<sup>A</sup>T<sub>E</sub>X file prepared by the author.